\documentclass[prx,aps,twocolumn,preprintnumbers,amsmath,amssymb,superscriptaddress]{revtex4-2}

\usepackage[utf8]{inputenc}
\usepackage[T1]{fontenc}
\usepackage{lmodern}
\usepackage{graphicx}
\usepackage{dcolumn}
\usepackage{bm}
\usepackage{textcomp}
\usepackage{ulem}
\usepackage{ifpdf}
\usepackage[squaren,Gray]{SIunits}
\usepackage{color}
\definecolor{red}{rgb}{1,0,0}
\definecolor{blue}{rgb}{0,0,1}
\definecolor{darkred}{rgb}{0.6,0,0}
\definecolor{darkblue}{rgb}{0,0,0.6}
\definecolor{darkgreen}{rgb}{0,0.5,0}
\definecolor{grey}{rgb}{0.6,0.6,0.6}
\definecolor{greyp}{rgb}{0.7,0.5,0.8} %%%% pinkish grey
\definecolor{violet}{rgb}{0.5,0,0.5}

\ifpdf
\usepackage{epstopdf}
\usepackage[pdftex,unicode,pdfstartview={FitH},pdfborder={0 0 0}]{hyperref}
\usepackage{hypcap}
\else
\usepackage[hypertex]{hyperref}
\fi
\hypersetup{
    bookmarksnumbered = true,
    colorlinks = true, linkcolor = darkblue,
    citecolor = darkblue, filecolor = darkblue,
    menucolor = darkblue, urlcolor = darkblue
}

\newcolumntype{R}{>{$\displaystyle}r<{$}}
\newcolumntype{C}{>{$\displaystyle}c<{$}}

\begin{document}
    
%\title{Coulomb-correlated states of moir\'e excitons\\ and charges on a MoSe$_2$/WS$_2$ semiconductor moir\'e lattice}

\title{Coulomb-correlated states of moir\'e excitons and charges\\ in a semiconductor moir\'e lattice}

\author{Borislav Polovnikov*}
\affiliation{Fakult\"at f\"ur Physik, Munich Quantum Center, and Center for NanoScience (CeNS), Ludwig-Maximilians-Universit\"at M\"unchen, Geschwister-Scholl-Platz 1, 80539 M\"unchen, Germany}
\affiliation{Max-Planck-Institut f\"ur Quantenoptik, Hans-Kopfermann-Straße 1, 85748 Garching bei M\"unchen, Germany}
\author{Johannes Scherzer*}
\affiliation{Fakult\"at f\"ur Physik, Munich Quantum Center, and Center for NanoScience (CeNS), Ludwig-Maximilians-Universit\"at M\"unchen, Geschwister-Scholl-Platz 1, 80539 M\"unchen, Germany}
\author{Subhradeep Misra}
\affiliation{Fakult\"at f\"ur Physik, Munich Quantum Center, and Center for NanoScience (CeNS), Ludwig-Maximilians-Universit\"at M\"unchen, Geschwister-Scholl-Platz 1, 80539 M\"unchen, Germany}
\author{Xin Huang}
\affiliation{Fakult\"at f\"ur Physik, Munich Quantum Center, and Center for NanoScience (CeNS), Ludwig-Maximilians-Universit\"at M\"unchen, Geschwister-Scholl-Platz 1, 80539 M\"unchen, Germany}
\affiliation{Beijing National Laboratory for Condensed Matter Physics, Institute of Physics, Chinese Academy of Sciences, Beijing 100190, People’s Republic of China}
\affiliation{School of Physical Sciences, CAS Key Laboratory of Vacuum Physics, University of Chinese Academy of Sciences, Beijing 100190, People’s Republic of China}
\author{Christian Mohl}
\affiliation{Fakult\"at f\"ur Physik, Munich Quantum Center, and Center for NanoScience (CeNS), Ludwig-Maximilians-Universit\"at M\"unchen, Geschwister-Scholl-Platz 1, 80539 M\"unchen, Germany}
\author{Zhijie Li}
\affiliation{Fakult\"at f\"ur Physik, Munich Quantum Center, and Center for NanoScience (CeNS), Ludwig-Maximilians-Universit\"at M\"unchen, Geschwister-Scholl-Platz 1, 80539 M\"unchen, Germany}
\author{Jonas G{\"o}ser}
\affiliation{Fakult\"at f\"ur Physik, Munich Quantum Center, and Center for NanoScience (CeNS), Ludwig-Maximilians-Universit\"at M\"unchen, Geschwister-Scholl-Platz 1, 80539 M\"unchen, Germany}
\author{Jonathan F\"orste}
\affiliation{Fakult\"at f\"ur Physik, Munich Quantum Center, and Center for NanoScience (CeNS), Ludwig-Maximilians-Universit\"at M\"unchen, Geschwister-Scholl-Platz 1, 80539 M\"unchen, Germany}
\author{Ismail Bilgin}
\affiliation{Fakult\"at f\"ur Physik, Munich Quantum Center, and Center for NanoScience (CeNS), Ludwig-Maximilians-Universit\"at M\"unchen, Geschwister-Scholl-Platz 1, 80539 M\"unchen, Germany}
\author{Kenji Watanabe}
\affiliation{Research Center for Functional Materials, National Institute for Materials Science, 1-1 Namiki, Tsukuba 305-0044, Japan}
\author{Takashi Taniguchi}
\affiliation{International Center for Materials Nanoarchitectonics, National Institute for Materials Science, 1-1 Namiki, Tsukuba 305-0044, Japan}
\author{Anvar~S.~Baimuratov}
\affiliation{Fakult\"at f\"ur Physik, Munich Quantum Center, and Center for NanoScience (CeNS), Ludwig-Maximilians-Universit\"at M\"unchen, Geschwister-Scholl-Platz 1, 80539 M\"unchen, Germany}
\author{Alexander H{\"o}gele}
\affiliation{Fakult\"at f\"ur Physik, Munich Quantum Center, and Center for NanoScience (CeNS), Ludwig-Maximilians-Universit\"at M\"unchen, Geschwister-Scholl-Platz 1, 80539 M\"unchen, Germany}
\affiliation{Munich Center for Quantum Science and Technology (MCQST), Schellingtra\ss{}e 4, 80799 M\"unchen, Germany}
       
\date{\today}

\begin{abstract}
Semiconductor moir\'e heterostructures exhibit rich correlation-induced many-body phenomena with signatures of emergent magnetism, Mott insulating states or generalized
Wigner crystals observed in optical spectroscopy by probing intralayer excitons~\cite{Regan2020,Tang2020,Xu2020}. However, as the staggered band alignment in the underlying WSe$_2$/WS$_2$ heterobilayer system separates photoexcited electrons and holes to form lowest-energy interlayer excitons \cite{Jin2019}, direct access to interactions between correlated charge states and ground state moir\'e excitons remained elusive. Here, we use MoSe$_2$/WS$_2$ heterostacks with type-I band alignment \cite{Tang2021,Tang2022} to probe Coulomb interactions of elementary charges with the ground and excited states of moir\'e excitons \cite{Alexeev2019}. In addition to positive and negative moir\'e trions \cite{LiuHeinz2021,BrotonsGisbert2021,Wang2021,Baek2021} reminiscent of their canonical counterparts in monolayer MoSe$_2$ \cite{Ross2013}, we reveal novel many-body states formed between moir\'e excitons and charges at fractional filling factors of the periodic moir\'e lattice. For both electrons and holes, these states exhibit doping-dependent Land\'e factors as a hallmark of correlation-induced magnetism \cite{Tang2020}, identifying the MoSe$_2$/WS$_2$ heterobilayer as a unique system for studies of correlated phenomena in ambipolar doping regimes.
\end{abstract}

\maketitle

Van der Waals heterostructures of twisted or lattice-mismatched two-dimensional transition metal dichalcogenides (TMDs) with electron, hole and exciton potentials laterally modulated by moir\'e effects \cite{WuTopo2017,Yu2017,Tong2017,WuHubbard2018,WuExciton2018} provide a rich platform for optical studies of correlation phenomena arising in flat bands \cite{Zhang2020_flat,LiCrommie2021} and spatially ordered states \cite{LiWang2021,Zhou2021}. Recent examples of many-body Hubbard model physics \cite{WuHubbard2018} include demonstrations of correlation-induced magnetism \cite{Tang2020}, Mott insulating states \cite{Regan2020,Shimazaki2020,Wang2020a,Huang2021,Ghiotto2021,Xu2020,LiMak2021mit}, quantum anomalous Hall effect \cite{LiMak2021} or Wigner crystals \cite{Regan2020,Zhou2021} of charges ordered on lattices with integer and fractional fillings \cite{LiWang2021} and stripes at half-filling factors \cite{Jin2021}. In heterobilayers (HBLs), related optical studies have focused on systems with type II band alignment as in WSe$_2$/WS$_2$ or MoSe$_2$/WSe$_2$, with lowest-energy states characterized by interlayer excitons. In the presence of doping, the photoluminescence (PL) of such systems is thus dominated by moir\'e trions \cite{LiuHeinz2021,BrotonsGisbert2021,Wang2021,Baek2021} with Land\'e factors inherited from neutral interlayer excitons, signifying vanishingly small perturbation of the exciton wave function by the surrounding elementary charges and absence of exciton-charge correlation effects. Optical signatures of correlated charge states, on the other hand, have been detected hundreds of meV above the interlayer exciton ground state via intralayer exciton absorption in WSe$_2$ or WS$_2$ monolayers \cite{Tang2020,Regan2020,Huang2021,LiWang2021,Xu2020}.

%%%%%%%%%%%%%%%%%%%%%%%%%%%%  FIG 1  %%%%%%%%%%%%%%%%%%%%%%%%%%%%%%
\begin{figure*}[t!]  
\includegraphics[scale=1.0]{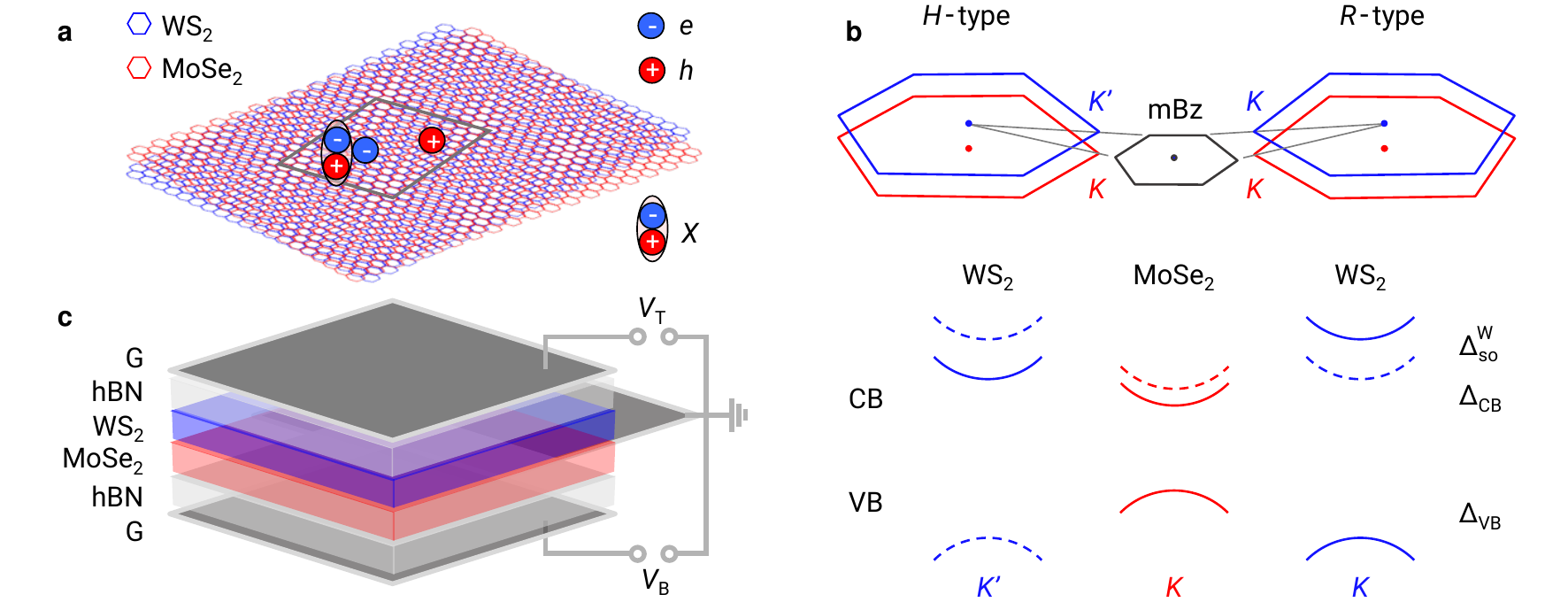}
\caption{\textbf{Schematics of MoSe$_2$/WS$_2$ moir\'e lattice, band structure and field-effect device.} \textbf{a}, Real-space moir\'e lattice of a twisted MoSe$_2$/WS$_2$ heterostructure with distinct locations of electrons ($e$), holes ($h$) and excitons ($X$) dictated by the spatially modulated moir\'e potential. \textbf{b}, Antiparallel $H$-type and parallel and $R$-type stackings give rise to $K-K'$ and $K-K$ alignment of MoSe$_2$ and WS$_2$ valleys and the formation of twist-angle dependent mini-Brillouin zone (mBz). Note the reversed ordering for $H$- and $R$-stacks of the spin-up (solid lines) and spin-down (dashed lines) polarized conduction sub-bands of WS$_2$ with spin-orbit splitting $\Delta_{\text{so}}^\text{W}$. Both stackings are in type-I band alignment with conduction and valence band offsets $\Delta_{\text{CB}}$ and $\Delta_{\text{VB}}$, respectively. \textbf{c}, Schematics of exfoliation-stacked MoSe$_2$/WS$_2$ HBL capped by hBN in a dual-gate field-effect device with $V_\text{T}$ and $V_\text{B}$ voltages applied to few-layer graphene top and bottom electrodes with respect to a grounded charge reservoir in contact with both monolayers.}
\label{fig1}
\end{figure*}
%%%%%%%%%%%%%%%%%%%%%%%%%%%%%%%%%%%%%%%%%%%%%%%%%%%%%%%%%%%%%%%%%%

In this work, we use a MoSe$_2$/WS$_2$ heterostructure to establish evidence for novel many-body states formed among ground-state moir\'e excitons and spatially ordered charges. As opposed to WSe$_2$/WS$_2$ HBLs, the MoSe$_2$/WS$_2$ system has received only little attention in the framework of correlated states in moir\'e flat bands, and instead raised some controversy regarding its band alignment \cite{Ceballos2015,Zhang2016,Alexeev2019,Meng2020,Zhang2020,Tang2021,Tang2022}. Notably, \textit{ab initio} calculations predicted a type-II band alignment, with the valence band (VB) maximum located at the $K$ valley of MoSe$_2$ and the conduction band (CB) minimum in the $K$ valley of WS$_2$ below the CB edge of MoSe$_2$ \cite{Gong2015_DFT}. This, however, was revised in recent work in charge-tunable MoSe$_2$/WS$_2$ heterostacks \cite{Tang2021,Tang2022, Zhang2016}, suggesting type-I band alignment, with the HBL band gap given by that of MoSe$_2$. Our study confirms this band alignment for twisted heterostacks near both high-symmetry configurations of antiparallel ($H$) and parallel ($R$) layer orientation with twist angles close to $180\degree$ and $0\degree$, respectively. With experimental access to ambipolar doping in field-effect devices, we observe distinct Coulomb-correlated states of positive and negative charges and moir\'e excitons, with signatures of charge order and correlation-induced magnetism at integer and fractional fillings of moir\'e unit cells.

Akin to WSe$_2$/WS$_2$ heterostacks, MoSe$_2$/WS$_2$ HBLs feature relatively large lattice incommensurability of $4\%$, which inhibits nanoscale and mesoscopic reconstruction ubiquitous in systems with small lattice mismatch as in MoS$_2$/WS$_2$ or MoSe$_2$/WSe$_2$ \cite{Weston2020,McGilly2020,Rosenberger2020,Halbertal2021,Shabani2021,Zhao2022}. Sizable lattice mismatch stabilizes moir\'e phenomena by locking HBLs rigidly in the canonical moir\'e geometry illustrated in Fig.~\ref{fig1}a. In this limit, the moir\'e pattern varies spatially through points of high-symmetry registries with gradual interconversion, giving rise to periodically modulated in-plane moir\'e potentials for electrons, holes and excitons \cite{WuTopo2017,WuHubbard2018,WuExciton2018} with distinct, energetically favored spatial positions within the moir\'e cell as illustrated in Fig.~\ref{fig1}a. The superlattice constant of the moir\'e unit cell depends sensitively on the rotation angle and attains its maximum at about $8$~nm~\cite{Kormanyos2015,Hermann2012} in both $H$- and $R$-limits. In reciprocal space (schematics in the top panel of Fig.~\ref{fig1}b), these two configurations correspond to $K-K'$ and $K-K$ alignments of the MoSe$_2$ and WS$_2$ valleys with twist-angle dependent mini-band dispersions in the mini-Brillouin zone \cite{Alexeev2019}. A key difference between $H$- and $R$-stackings is the ordering of spin-polarized conduction sub-bands of WS$_2$, as illustrated in the bottom panel of Fig.~\ref{fig1}b.

%%%%%%%%%%%%%%%% up to here

%%%%%%%%%%%%%%%%%%%%%%%%  FIG 2  %%%%%%%%%%%%%%%%%%%%%%%%%%%%%%%%%%%
\begin{figure*}[ht!]
\includegraphics[scale=1.0]{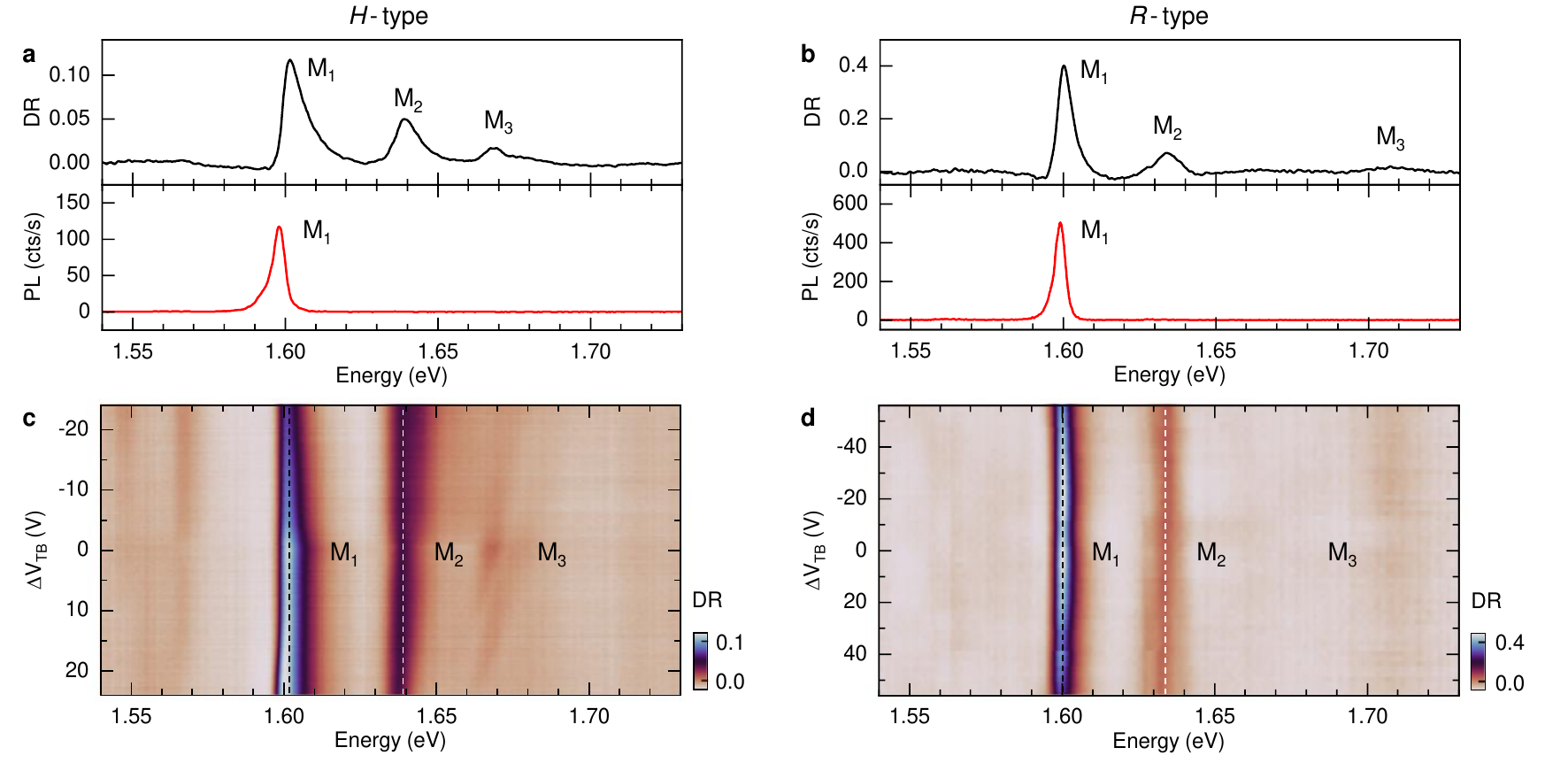}
\caption{\textbf{Moir\'e excitons in charge-neutral MoSe$_2$/WS$_2$ heterostacks.} \textbf{a} and \textbf{b}, DR (top panel) and PL (bottom panel) spectra of $H$- and $R$-type heterostacks at charge neutrality with moir\'e exciton peaks $M_1$, $M_2$ and $M_3$; $M_1$ is the moir\'e exciton ground state and thus dominates PL due to population relaxation. \textbf{c} and \textbf{d}, Dispersion of moir\'e peaks in perpendicular electric field (proportional to $\Delta V_\text{TB}$) recorded in DR for $H$- and $R$-type heterostacks, respectively. The dashed lines indicate zero dispersion, signifying marginal interlayer character of the lowest two moir\'e exciton states $M_1$ and $M_2$. All data were recorded at $3.2$~K.}
\label{fig2}
\end{figure*}
%%%%%%%%%%%%%%%%%%%%%%%%%%%%%%%%%%%%%%%%%%%%%%%%%%%%%%%%%%%%%%%%%%%%%%%%%%%

To probe the charging behavior of MoSe$_2$/WS$_2$ HBLs at both high-symmetry stackings, we fabricated charge-tunable van der Waals devices illustrated in Fig.~\ref{fig1}c. In all devices, the heterostacks were encapsulated in hexagonal boron nitride (hBN) and sandwiched between top and bottom few-layer graphene electrodes. Two devices were assembled from MoSe$_2$ and WS$_2$ monolayers synthesized by chemical vapor deposition (CVD) with triangular shapes facilitating relative orientation (see Extended Data Fig.~1), and two more devices were fabricated from monolayers exfoliated from native crystals. From optical alignment of crystallographically terminated crystal edges we aimed at rotational twist angles around $0\degree$ ($180\degree$) in all samples. The dual-gate layout of the field-effect device allows us to subject the MoSe$_2$/WS$_2$ HBL to perpendicular electric field by an imbalanced tuning of the top and bottom gate voltages $V_\text{T}$ and $V_\text{B}$ as $\Delta V_\text{TB}=V_\text{T}-V_\text{B}$ with respect to the grounded reservoir in contact with both MoSe$_2$ and WS$_2$ monolayers, or to vary the doping level by balancing both gates and tuning them simultaneously against the ground as gate voltage $V_\text{G}=V_\text{T}=V_\text{B}$. The former regime probes the static out-of-plane exciton dipole moment via the Stark effect, and the latter allows to tune the Fermi level through the spatially modulated electron and hole potentials of the moir\'e heterostructure.

We first focus on the charge-neutral regime of moir\'e excitons in MoSe$_2$/WS$_2$. The top panels of Fig.~\ref{fig2} show cryogenic differential reflection DR (defined as $(\textrm{R}_0-\textrm{R})/\textrm{R}_0$ with reflection $\textrm{R}$ on the HBL region and the reference $\textrm{R}_0$ away from it), and PL spectra of $H$-type (Fig.~\ref{fig2}a) and $R$-type (Fig.~\ref{fig2}b) heterostacks tuned to charge neutrality. In the spectral range between $1.54$ and $1.73$~eV, the peaks in DR reflect the multiplicity of moir\'e excitons in twisted MoSe$_2$/WS$_2$ HBLs \cite{Alexeev2019,Tang2021,Tang2022}. In the DR spectra of both stackings (top panels), the $M_1$ peak at $1.60$~eV, about $30$~meV below the transition energy of the fundamental exciton in monolayer MoSe$_2$ at $1.63$~eV in our samples, exhibits largest oscillator strength. The consecutive moir\'e peak $M_2$ is observed at $38$ and $35$~meV blueshift from $M_1$ in $H$- and $R$-type spectra, and the respective blueshifts of $M_3$ are $65$ and $110$~meV. The PL spectra of both stacks (bottom panels) feature only the emission from the ground state moir\'e exciton $M_1$ due to efficient population relaxation from higher-energy states. 

The energetic ordering of moir\'e exciton peaks detected in DR is consistent with type-I band alignment of both $H$- and $R$-type heterostacks with CB and VB offsets of a few tens and a few hundreds of meV, respectively \cite{Tang2021,Tang2022}. The large VB offset manifests pure MoSe$_2$ VB-character for all moir\'e excitons. The one order of magnitude smaller CB offset in turn implies pure MoSe$_2$ CB-character only for the lowest-energy moir\'e states $M_1$ and $M_2$, and a hybrid interlayer character for the energetically highest state $M_3$. With this notion, and taking into account that interlayer hybridization favors spin-like bands, the DR spectra in Fig.~\ref{fig2}a and b are readily explained: the first two moir\'e states $M_1$ and $M_2$ correspond to the renormalized intralayer exciton of MoSe$_2$ and its first Umklapp peak due to the moir\'e potential. Their energy separation is determined by the relative twist angle between the two monolayers and is not sensitive to interlayer coupling. The moir\'e exciton state $M_3$, in contrast, is formed by VB states of MoSe$_2$ and CB states of WS$_2$ with spin-like character, and the difference in the blueshifts between the respective $M_3$ and $M_1$ peaks is an immediate consequence of the reversed ordering of spin-polarized WS$_2$ sub-bands in $H$- and $R$-type stackings. For ideal $H$- and $R$-type alignments and in the limit of vanishing interlayer coupling illustrated in Fig.~\ref{fig1}b, the difference in the blueshifts would be given by $\Delta_\text{so}^\text{W}$, the spin-orbit splitting in the CB of WS$_2$. In the DR spectra of Fig.~\ref{fig2}a and b, the actual difference of $45$~meV between the blueshifts of $M_1$ and $M_3$ moir\'e peaks implicitly accounts for finite twist and interlayer coupling. 

The degree of interlayer character of the moir\'e peaks can be probed experimentally via their dispersion in perpendicular electric field, as obtained from the evolution of the DR spectra with $\Delta V_\text{TB}$ in Fig.~\ref{fig2}c and d for $H$- and $R$-heterostacks. The linear slopes of the moir\'e peak dispersions reflect the first-order Stark effect proportional to their out-of-plane electrostatic dipole moment, which in turn is a measure of electron delocalization over the two layers with respect to the hole in the MoSe$_2$ layer. In both stackings, $M_1$ and $M_2$ exhibit vanishing dispersions in electric field, confirming their intralayer MoSe$_2$ character. Consistently, the moir\'e peak $M_3$ exhibits finite Stark slopes in both $H$- and $R$-type heterostacks due to its hybrid interlayer exciton character (see Extended Data Fig.~2 for a quantitative analysis).  

%%%%%%%%%%%%%%%%%%%%%%%%  FIG 3  %%%%%%%%%%%%%%%%%%%%%%%%%%%%%%%%%%%
\begin{figure*}[ht!]
\includegraphics[scale=1.0]{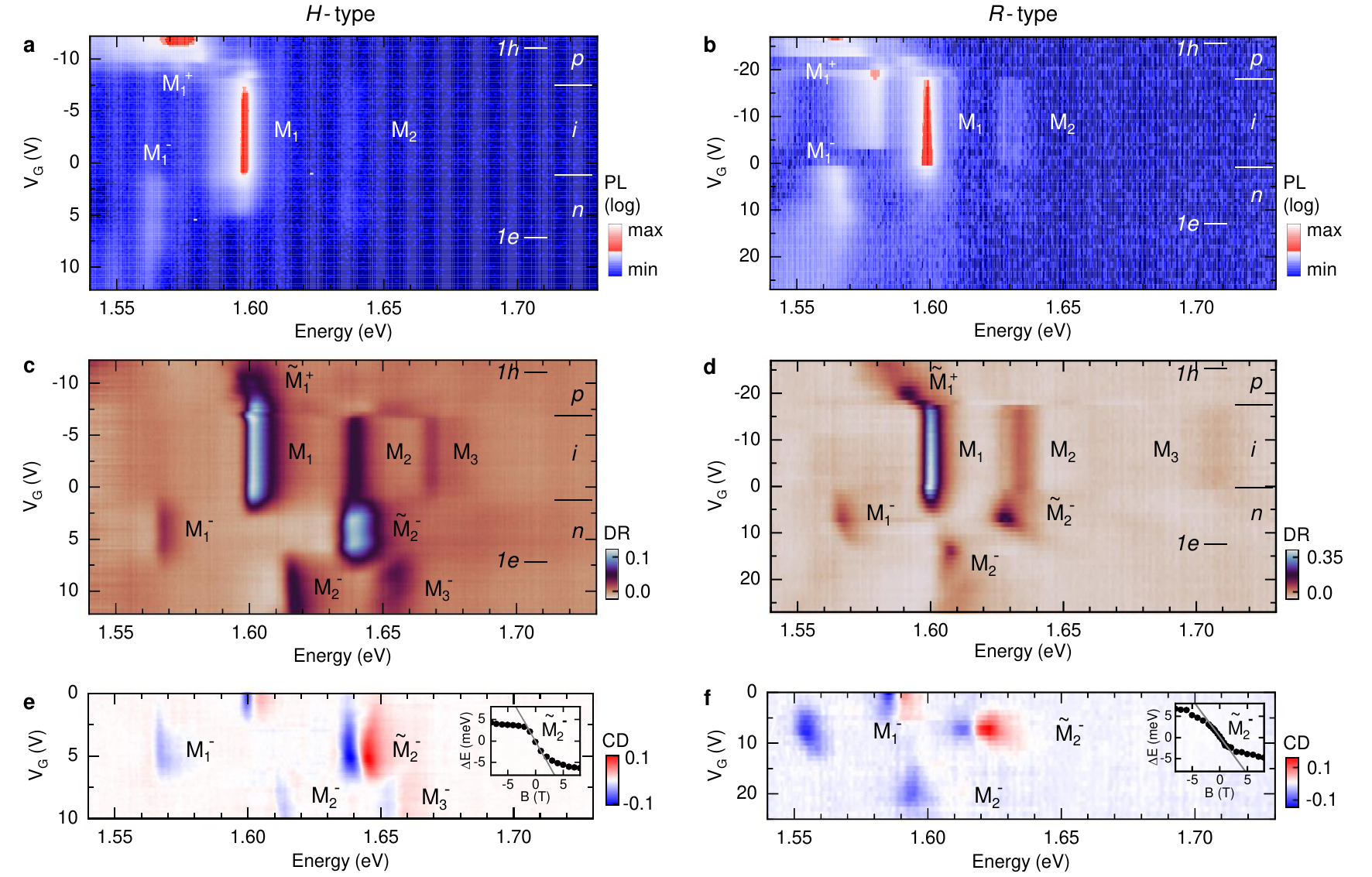}
\caption{\textbf{Ambipolar field-effect doping regimes of MoSe$_2$/WS$_2$ heterostacks.} \textbf{a} and \textbf{b}, Evolution of PL in $H$- and $R$-type stacks, respectively, as the charge carrier density is tuned with the gate voltage $V_\text{G}$ from the positive ($p$) through the intrinsic ($i$) into the negative ($n$) doping regime ($1e$ and $1h$ indicate filling factors of $1$ electron and hole per moir\'e unit cell). Hole and electron doping converts the neutral moir\'e exciton peak $M_1$ into the respective canonical positive and negative trions $M^{+}_1$ and $M^{-}_1$ with binding energies of $24$ and $33$~meV in $H$-type stacks, and equal binding energies of $35$~meV for both charge carriers in the $R$-type spectra. In both stackings, $M_2$ is present as weak hot-luminescence; the peak between $M_1$ and $M^{+}_1$ in the $R$-type charging diagram is not labelled due to unclear origin. \textbf{c} and \textbf{d}, Corresponding DR charging diagrams for $H$- and $R$-stacks, respectively. Note the emergence of additional strong peaks $\tilde{M}^{+}_1$ and $\tilde{M}^{-}_2$ formed by moir\'e excitons $M_1$ and $M_2$ upon charging with holes and electrons. \textbf{e} and \textbf{f}, Corresponding circular dichroism (CD) spectra at a magnetic field of $-8$~T. The peaks $M_1^-$ and $M_2^-$ show strong valley polarization reminiscent of MoSe$_2$ monolayer trions, whereas $\tilde{M}_2^-$ shows a Zeeman splitting that is strongly enhanced compared to that of the neutral moir\'e exciton $M_1$. The insets show  nonlinear valley Zeeman splittings of $\tilde{M}_2^-$ as a function of magnetic field. Note that the CD data for the $R$-type sample in \textbf{f} were acquired at a different position than the data shown in \textbf{b} and \textbf{d} with an energy offset in the peak energies due to a different dielectric environment. All data were recorded at $3.2$~K.}
\label{fig3}
\end{figure*}
%%%%%%%%%%%%%%%%%%%%%%%%%%%%%%%%%%%%%%%%%%%%%%%%%%%%%%%%%%%%%%%%%%%%%%

Having established the nearly-pure MoSe$_2$ intralayer character for the ground and first excited moir\'e exciton states $M_1$ and $M_2$ in both stackings, we focus on their response to charge-carrier doping in Fig.~\ref{fig3}. With increasing gate voltages $V_\text{G}$, the charging characteristics of PL (Fig.~\ref{fig3}a and b) and DR (Fig.~\ref{fig3}c and d) exhibit for both stackings transitions from the positive ($p$) through the charge-neutral intrinsic ($i$) to the negative ($n$) doping regimes, reaching on the positive (negative) side into doping levels beyond one hole (electron) per moir\'e cell. In both stackings, the PL in the neutral regime is dominated by the ground state peak $M_1$, accompanied by very weak hot-luminescence from the excited moir\'e state $M_2$. Another unlabelled feature appears $20$~meV below $M_1$ in the $R$-type sample that vanishes towards charge-neutrality and has no counterpart in the charging diagram of the $H$-stack. At highest $p$-doping levels, reached in PL due to non-resonant laser excitation that generates a surplus of holes, the positive trion peak $M_1^+$ emerges at the expense of its neutral counterpart $M_1$, with binding energies of $24$ and $35$~meV in $H$- and $R$-type stacks. Upon electron doping, $M_1$ converts into the negative trion $M_1^-$ with similar binding energies of $33$ and $35$~meV in $H$- and $R$-stacks. These positive and negative trions correspond to canonical three-particle complexes with mutual on-site Coulomb interactions, reminiscent of trions in monolayer MoSe$_2$ with nearly identical binding energies \cite{Ross2013} in the low-density regimes of Fermi polarons \cite{Sidler2017}. In both stackings, the hot-luminescence peak $M_2$ exhibits no obvious signatures on the $p$-doping side, and a very weak feature with a small energy shift from $M_2$ upon $n$-doping.

%%%%%%%%%%%%%%%%%%%%%%%%  FIG 4  %%%%%%%%%%%%%%%%%%%%%%%%%%%%%%%%%%%
\begin{figure*}[ht!]
\includegraphics[scale=1.0]{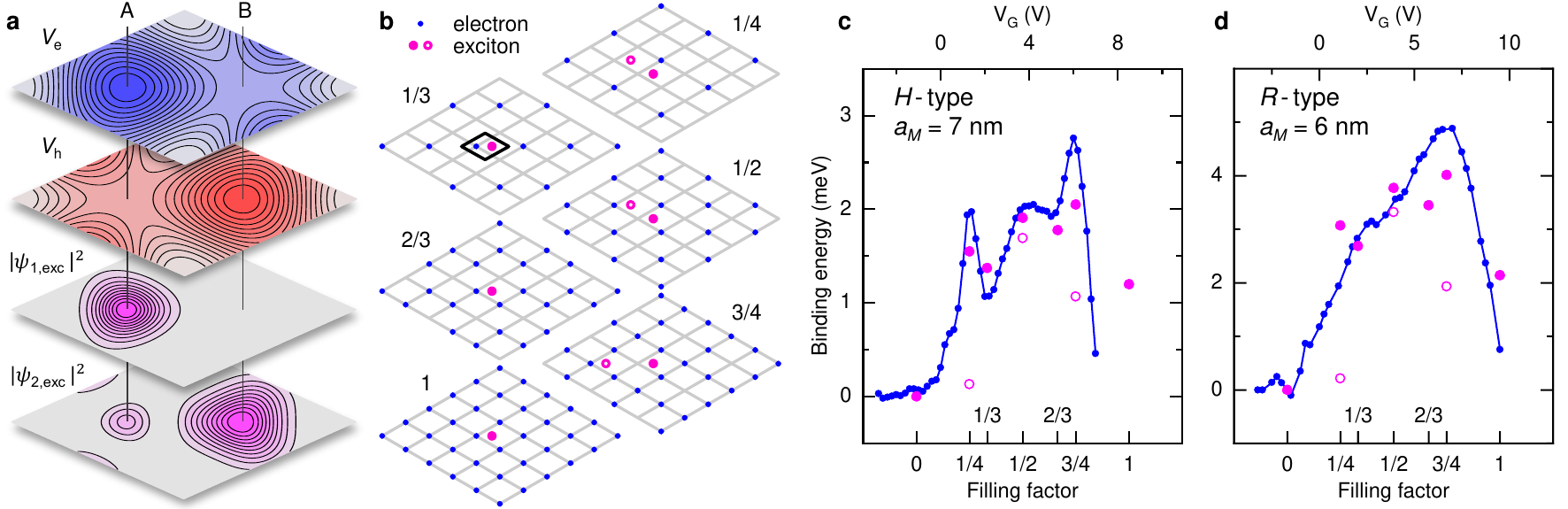}
\caption{\textbf{Interactions of moir\'e excitons with a charge lattice at integer and fractional fillings.} \textbf{a}, From top to bottom: electron and hole potentials $V_e$ and $V_h$ and probability distributions of moir\'e excitons $M_1$ and $M_2$ with wave functions $\psi_{1,\mathrm{exc}}$ and $\psi_{2,\mathrm{exc}}$ within one moir\'e unit cell. Minimum-energy electron and hole localization sites are denoted by sites A and B. \textbf{b}, Spatial positions of one exciton (magenta circles) interacting with electrons (blue circles) on the moir\'e lattice at mean fractional fillings of up to one charge per cell (note the two inequivalent exciton positions denoted by open and closed circles for even-denominator fractional fillings). \textbf{c} and \textbf{d}, Experimental (blue data) and theoretical (open and closed magenta points) binding energy of the state $\tilde{M}_2^-$ in $H$- and $R$-type heterostacks, respectively, evaluated as its redshift from $M_2$ as a function of electron filling factor.}
\label{fig4}
\end{figure*}
%%%%%%%%%%%%%%%%%%%%%%%%%%%%%%%%%%%%%%%%%%%%%%%%%%%%%%%%%%%%%%%%%%%%%%%%

Remarkably, the corresponding feature, labelled as $\tilde{M}_2^-$ in Fig.~\ref{fig3}c and d, is very prominent in DR throughout the $n$-regime below one electron per moir\'e cell, before it abruptly converts into the redshifted peak $M_2^-$ upon further electron-doping. This transition coincides with the emergence of $M_3^-$ in the $H$-type DR charging diagram in Fig.~\ref{fig3}c. Further inspection of the charging behavior reveals a surprising asymmetry in the responses to hole and electron doping. On the hole-doping side, the peak $M_1$ undergoes a transition to $M_1^+$ through consecutive step-like states labelled as $\tilde{M}_1^+$, whereas $M_2$ and $M_3$ disappear abruptly. This charging behavior is reversed on the electron-doping side: the transition from $M_1$ to $M_1^-$ is abrupt, whereas $M_2$ evolves gradually into $\tilde{M}_2^-$ before jumping abruptly to $M_2^-$. Such contrasting electron and hole charging sequences were observed consistently in all samples of our study, with only marginal variations in the energy shifts between charge-dependent features (see Extended Data Fig.~3).

The nature of the charged transitions in both doping regimes can be elucidated by probing their response to magnetic fields. We focus on the electron-doping side, and show in Fig.~\ref{fig3}e and f the degree of circular dichroism (CD) for $H$- and $R$-stacks, determined as $\text{CD} = \text{DR}^+ - \text{DR}^-$ from the difference of $\sigma^+$ and $\sigma^-$ polarized DR spectra in a perpendicular magnetic field of $-8$~T. According to optical selection rules, positive (red) and negative (blue) CD values signify the degree of valley polarization via $\sigma^+$ and $\sigma^-$ circularly polarized exciton transitions, as well as the magnitude of the valley Zeeman splitting by the energy difference between the maximum and minimum CD peak values. In both stackings, the peaks $M_1^-$ and $M_2^-$ feature almost complete valley polarization, just like Fermi polarons in monolayer MoSe$_2$ \cite{Back2017}. The peak $\tilde{M}_2^-$, on the other hand, is unpolarized and shows a doping-dependent Zeeman splitting with nonlinear dependence on the magnetic field as in the insets of Fig.~\ref{fig3}e and f. At magnetic field values below $2$~T, the corresponding $g$-factor of  $\tilde{M}_2^-$ is enhanced by nearly ten times, as referenced to its asymptotic value of $-4$ at higher fields. We observed qualitatively similar behavior for $\tilde{M}_1^+$ in the $p$-doped regime (see Extended Data Fig.~5). Both the doping-dependence of the the $g$-factors and their nonlinear evolution with magnetic field are indicative of correlation-induced magnetism \cite{Tang2020} of $\tilde{M}_2^-$ and $\tilde{M}_1^+$ states as probes of emergent charge order at integer and fractional fillings of the moir\'e unit cell.

First, we focus on the limit of one charge per moir\'e cell. The spatially varying intracell electron and hole moir\'e potentials $V_e$ and $V_h$, calculated using the continuum model \cite{WuTopo2017} and shown as the two top layers of Fig.~\ref{fig4}a, define charge localization sites A and B at the respective potential minima (blue- and red-most spots in the surface plots of Fig.~\ref{fig4}a for electrons and holes). The two lowest moir\'e exciton states $M_1$ and $M_2$, with probability distributions shown in the two bottom layers of Fig.~\ref{fig4}a as obtained from continuum models \cite{WuHubbard2018,WuExciton2018}, form two distinct spatially co-localized pairs with electrons and holes: the ground state moir\'e exciton $M_1$ is co-localized with the electron on site A, whereas the excited moir\'e state $M_2$ is co-localized with the hole on site B. These two configurations are reminiscent of canonical negative and positive trions in monolayer MoSe$_2$. Consistently, as the charge carriers at such low doping levels are confined to the MoSe$_2$ layer of the heterostack, the binding energy of the negative trion $M_1^-$ in Fig.~\ref{fig3} is comparable with the value in MoSe$_2$ monolayer \cite{Ross2013} irrespective of the stacking type (see also Extended Data Fig.~4). 

At doping levels exceeding one charge per moir\'e cell, Coulomb repulsion prevents double occupancy of the same site. Thus, in the $n$-doped regime, the second electron will avoid the site A and instead populate the site B to form the negative trion $M_2^-$ co-localized with the moir\'e exciton $M_2$. In the $p$-doped regime, the second hole will accordingly co-localize with $M_1$ at site A to form the positive trion $M_1^+$ observed in PL in Fig.~\ref{SI_fig3}. Consistently, the $M_1^+$ states of both stackings have binding energies comparable to positive trions in monolayer MoSe$_2$. The binding energies of $M_2^-$, on the other hand, are reduced in accord with partial delocalization of the moir\'e exciton wave function between the sites A and B as illustrated in the bottom layer of Fig.~\ref{fig4}b. 

For doping levels below integer filling per moir\'e cell, the common picture of trion formation between one charge and one exciton \cite{Ross2013} clearly breaks down. The alternative notion of polaron formation between an exciton and a Fermi sea of charges \cite{Sidler2017} seems also inapplicable given the spatial ordering of electrons and holes by the in-plane moir\'e potential. In this regime, $\tilde{M}_1^+$ and $\tilde{M}_2^-$ rather reflect many-body correlated states formed by Coulomb interactions between ordered charges and moir\'e excitons $M_1$ and $M_2$, with signatures of correlation-induced magnetism discussed above. We obtain additional insight into the nature of these states from theoretical modeling. To this end, we consider an exciton pinned on one site of the moir\'e unit cell and subjected to Coulomb interactions with charge lattices of varying geometry for different fractional fillings as indicated in Fig.~\ref{fig4}b for the case of electron doping. Using the variational approach (see Methods for details), we calculate the change in the exciton binding energy in the presence of an ordered charge lattice, and compare it in Fig.~\ref{fig4}c and d to the redshift of $\tilde{M}_2^-$ from the respective $M_2$ moir\'e peak. The quantitative agreement between experiment (blue data points) and theory (magenta points) is compelling: as the filling factor (bottom axis in the theory model) is increased from zero to one electron per moir\'e cell with increasing gate voltage (top axis of actual experiments), the binding energy varies from zero up to a maximum of $3$ and $5$~meV in $H$- and $R$-type stacks, respectively, providing an estimate for the energy scale of many-body interactions between excitons and charges ordered on the underlying moir\'e lattice. 

With this understanding of the rich phenomena observed in MoSe$_2$/WS$_2$ HBLs upon ambilopar doping, our findings identify the system as a unique material platform for studies of correlated phenomena induced by charge ordering in both electron- and hole-doped regimes. Given the rich variety of theoretically predicted phases in related settings \cite{WuHubbard2018,Pan2020,Devakul2021,Zhang2021,Zang2021,Hu2021}, the specific heterostructure stands out by providing access to contrasting electron and hole moir\'e potentials, effective masses and angular momenta upon doping in relation to the same set of moir\'e excitons. Our insight into the nature of the emergent Coulomb-correlated states at integer and fractional moir\'e cell fillings with doping-dependent magnetism provide compelling motivation for future experimental and theoretical work on many-body phenomena in MoSe$_2$/WS$_2$ heterostacks.

%%%%%%%%%%%%%%%%%%%%%%%%%%%%%%%%%%%%%%%%%%%%%%%%%%%%%%%%%%%%%%%%%%%%%%%%%%%%%%%%%%
%\clearpage
\section*{Methods}
\noindent \textbf{Device fabrication:} MLs of MoSe$_2$ and WS$_2$ were either mechanically exfoliated from bulk crystals (HQ Graphene) or obtained from in-house CVD synthesis. Thin flakes of hBN were exfoliated from bulk crystals (NIMS). Devices from hBN-encapsulated MoSe$_2$/WS$_2$ HBLs were prepared by dry exfoliation-transfer with alignment close to $0\degree$ ($R$-type) and $180\degree$ ($H$-type) by selecting straight crystal edges.
%%%%%%%%%%%%%%%%%%%%%%%%%%%%%%%%%%%%%%%%%%%%%%%%%%%%%%%%%%%%%%%%%%%%%%%%%%%%%%%%%%
\vspace{8pt}
\\    
\noindent \textbf{Optical spectroscopy:} Cryogenic PL and DR spectroscopy was conducted using home-built confocal microscopes in back-scattering geometry. The samples were loaded into a closed-cycle cryostat (attocube systems, attoDRY1000) with a base temperature of $3.2$~K or a dilution refrigerator (Leiden Cryogenics) operated at $4$~K. Both cryogenic systems were equipped with a superconducting magnet providing magnetic fields of up to $\pm 9$~T in Faraday configuration. Piezo-stepping and scanning units (attocube systems, ANPxyz and ANSxy100) were used for sample positioning with respect to a low-temperature apochromatic objective (attocube systems). A wavelength-tunable Ti:sapphire laser (Coherent, Mira) in continuous-wave mode and laser diodes were used to excite PL. For DR measurements, a stabilized Tungsten-Halogen lamp (Thorlabs, SLS201L) and supercontinuum lasers (NKT Photonics, SuperK Extreme and SuperK Varia) were used as broadband light sources. The PL or reflection signal were spectrally dispersed by monochromators (Roper Scientific, Acton SP2500 or Acton SpectraPro 300i with a 300 grooves/mm grating) and detected by liquid nitrogen or Peltier cooled charge-coupled devices (Roper Scientific, Spec-10:100BR or Andor, iDus 416). A set of linear polarizers (Thorlabs, LPVIS), half- and quarter-waveplates (B. Halle, $310-1100$~nm achromatic) mounted on piezo-rotators (attocube systems, ANR240) were used to control the polarization in excitation and detection. %The DR spectra were obtained by normalizing the reflected spectra from the HBL region ($R$) to that from the sample region without MoSe$_2$ and WS$_2$ layers ($R_0$) as $\textrm{DR} = (R-R_0)/R_0$. 
%%%%%%%%%%%%%%%%%%%%%%%%%%%%%%%%%%%%%%%%%%%%%%%%%%%%%%%%%%%%%%%%%%%%%%%%%%%%%%%%%%%
\vspace{8pt}
\\
\noindent\textbf{Electron and hole moir\'e potentials and moir\'e exciton wave functions:}
The two lowest moir\'e exciton states in type-I MoSe$_2$/WS$_2$ HBLs of $H$- and $R$-type can be considered as intralayer moir\'e states in the MoSe$_2$ layer. To illustrate the probability distribution for the moir\'e exciton states $M_1$ and $M_2$ shown in the two bottom panels of Fig.~\ref{fig4}a, we assume that the electron and hole potential landscapes vary periodically \cite{WuTopo2017,WuHubbard2018,WuExciton2018}:
\begin{equation}
  \Delta_q = \sum_{j=1}^6 V^{(q)}_j \exp(\text{i}\mathbf{b}_j \mathbf{r}), \nonumber
\end{equation}
where $q= \text{e},\text{h}$; $\mathbf{b}_j$ are the first-star reciprocal lattice vectors of the moir\'e pattern, and:
\begin{equation}
  V^{(q)}_{1,2,3} =  V^{(q)*}_{2,4,6} = V_q \exp(\text{i}\psi_q). \nonumber
\end{equation}
Taking into account that the exciton binding energy is much larger than the potential amplitudes $V_\text{e}$ and $V_\text{h}$, we approximate the potential modulation for the exciton center-of-mass as:
\begin{equation}
  \Delta_\text{exc} =  \Delta_\text{e} +  \Delta_\text{h}. \nonumber
\end{equation}
Within this framework we calculate the wave functions of moir\'e excitons with the following parameters: $(V_\text{e},\psi_\text{e}) = (14$~meV, $40^\circ$) and $(V_\text{h},\psi_\text{h}) = (7$~meV, $-40^\circ$), the moir\'e superlattice constant is $8$~nm and the exciton mass is $1.44~m_0$. Using these parameters we plot Fig.~\ref{fig4}a, illustrating that the lowest-energy moir\'e exciton state $M_1$ is localized at the minimum of the electron potential, whereas the second moir\'e exciton $M_2$ is mainly localized at the minimum of the hole potential.
\vspace{8pt}
\\
\noindent \textbf{Coulomb-interaction energies:}
To calculate the interaction energy (binding energy) of moir\'e excitons with electrons/holes ordered on a lattice we assume that the exciton is confined in one moir\'e cell and interacts with surrounding electrons/holes on a lattice as illustrated in Fig.~\ref{fig4}b. For each fractional filling we consider a lattice of elementary point charges with different spatially ordered patterns. Due to the lattice symmetry at a given fractional filling, distinct inequivalent positions of excitons are to be considered (empty and filled circles in Fig.~\ref{fig4}b) for $1/4$, $1/2$, and $3/4$ fillings.

We further assume that the main contribution to the binding energy stems from charge-induced modification of the electron-hole relative motion $\bm \rho \equiv (\rho,\theta) = \mathbf{r}_\text{e} - \mathbf{r}_\text{h}$, where $\mathbf{r}_\text{e(h)}$ are the coordinates of the electron and hole forming the exciton. The corresponding Schr\"odinger equation takes the form:
\begin{equation}
  -\frac{\hbar^2}{2\mu} \Delta \varphi(\bm\rho) + [V_\text{RK}(\rho) + V(\bm\rho)] \varphi(\bm\rho) = E \varphi(\bm\rho), \nonumber
\end{equation}
where $E$ is the exciton energy, $\mu = m_\text{e} m_\text{h}/(m_\text{e}+m_\text{h})$ is the reduced exciton mass, $m_\text{e}$ and $m_\text{h}$ are the electron and hole effective masses, and the Rytova--Keldysh potential~\cite{Rytova1967,Keldysh1979} of the electron-hole attraction is given by:
\begin{equation}
  V_\text{RK}(\rho) = -\frac{\pi e^2}{2 \varepsilon \rho_0} \left[ H_0 \left(\frac{\rho}{\rho_0}\right) - Y_0 \left(\frac{\rho}{\rho_0}\right) \right]. \nonumber
\end{equation}
Here, $e$ is the electron charge, $\rho_0$ is the screening length, $\varepsilon$ is the effective dielectric constant, and $H_0 (x)$ and $Y_0 (x)$ are Struve and Neumann functions. 

The interaction of the exciton with the charge lattice is described by the Coulomb sum:
\begin{equation}
V(\bm\rho) = \pm \frac{e^2}{\varepsilon} \sum_\mathbf{n}
                                \left[
                                  \frac{1}{|\beta_\text{e} \bm\rho + \mathbf{n}|}
                                  -\frac{1}{|\beta_\text{h} \bm\rho - \mathbf{n}|}
                                \right],
                                \nonumber
\end{equation}
where the plus and minus signs correspond to positive and negative elementary charges, $\beta_\text{e} = m_\text{e}/(m_\text{e}+m_\text{h})$, $\beta_\text{h} = m_\text{h}/(m_\text{e}+m_\text{h})$, and $\mathbf{n}$ are the coordinates of electrons/holes on the lattice. The two terms in the brackets determine the interaction of the charge lattice with the hole and the electron that constitute the exciton. 

To determine the binding energy of the state, we calculate the free exciton energy $E_X$ to obtain:
\begin{equation}
  E_\text{B} = E_X - E. \nonumber
\end{equation}
To calculate $E_X$, we set $V(\bm \rho)=0$, and use in the calculations of both $E_X$ and $E$ 2D hydrogen-like wave functions with the Bohr radius as variational parameter~\cite{Chernikov2014,Courtade2017,Semina2019} and the basis of six functions~\cite{Yang1991} with quantum numbers $(n,l) = (1,0),(2,0),(2,\pm1),(4,\pm3)$ to take into account polarization effects on the exciton relative motion. Due to the lower rotational symmetry of the potential $V(\bm\rho)$, we also include hydrogen-like wave functions with angular momenta $l = \pm 1, \pm 3$. The explicit expression for the trial function is:
\begin{equation}
  \varphi(\rho,\theta) = e^{-\alpha \rho}
                          + \zeta \rho e^{-\beta \rho}
                           + \eta \rho e^{-\gamma \rho} \cos\theta
                            + \xi \rho^3 e^{-\delta \rho} \cos3\theta.
                             \nonumber
\end{equation}
We solve the minimization problem numerically for seven parameters $(\alpha, \beta, \gamma, \delta, \zeta, \eta, \xi)$ using MATLAB R2017B and experimental material parameters of MoSe$_2$ monolayers~\cite{Goryca2019}: $m_\text{e} = 0.84 m_0$, $m_\text{h} = 0.6m_0$, $\varepsilon = 4.4$, $\rho_0 = 0.89$~nm. The only fitting parameter for comparison between experimental data and theoretical model results is the moir\'e superlattice constant. In the main text we fit the data with $7$ and $6$~nm for $H$- and $R$-type MoSe$_2$/WS$_2$ HBLs, respectively, corresponding to $178.6^\circ$ and $2.2^\circ$ twist angles.
%The dependence of the binding energy on the superlattice constant is shown in Extended Data Fig.~\ref{figTheory}. The trion binding energies strongly depend on the ratio of electron and hole effective masses. Since the electron mass is larger than that of the hole, negative doping results only in positive binding energies. In contrast, for some fractional fillings in the positive doping regime, the energies of the exciton interacting with the hole lattice is blue-shifted with respect to the free exciton. This happens when the polarization effects cannot overcome the energy increase, due to the repulsive interaction between the exciton hole and the positively charged lattice. On the other hand, for negative doping the polarization effects and the attractive interaction between the exciton hole and the electron lattice interfere constructively. It should be noted that for given material parameters the exciton ``core'' is negatively charged, oppositely to the hydrogen problem.
%%%%%%%%%%%%%%%%%%%%%%%%%%%%%%%%%%%%%%%%%%%%%%%%%%%%%%%%%%%%%%%%%%%%%%%%%%%%%%%%%%%
%%%%%%%%%%%%%%%%%%%%%%%%%%%%%%%%%%%%%%%%%%%%%%%%%%%%%%%%%%%%%%%%%%%%%%%%%%%%%%%%%%%
\vspace{8pt}
\\
\noindent \textbf{Acknowledgements:}\\
This research was funded by the European Research Council (ERC) under the Grant Agreement No.~772195 as well as the Deutsche Forschungsgemeinschaft (DFG, German Research Foundation) within the Priority Programme SPP~2244 2DMP and the Germany's Excellence Strategy EXC-2111-390814868. B.\,P acknowledges funding by IMPRS-QST. I.\,B. acknowledges support from the Alexander von Humboldt Foundation. X.\,H. and A.\,S.\,B. received funding from the European Union's Framework Programme for Research and Innovation Horizon 2020 (2014--2020) under the Marie Sk{\l}odowska-Curie Grant Agreement No.~754388 (LMUResearchFellows) and from LMUexcellent, funded by the Federal Ministry of Education and Research (BMBF) and the Free State of Bavaria under the Excellence Strategy of the German Federal Government and the L{\"a}nder. Z.\,Li. was supported by the China Scholarship Council (CSC), No. 201808140196. K.W. and T.T. acknowledge support from JSPS KAKENHI (Grant Numbers 19H05790, 20H00354 and 21H05233).
%%%%%%%%%%%%%%%%%%%%%%%%%%%%%%%%%%%%%%%%%%%%%%%%%%%%%%%%%%%%%%%%%%%%%%%%%%%%%%%%%%%
\vspace{8pt}
\\
%%%%%%%%%%%%%%%%%%%%%%%%%%%%%%%%%%%%%%%%%%%%%%%%%%%%%%%%%%%%%%%%%%%%%%%%%%%%%%%%%%%
\noindent \textbf{Contributions:}\\
J.\,G., Z.\,Li. and I.\,B. synthesized monolayers by CVD crystal growth. K.\,W. and T.\,T. provided high-quality hBN crystals. X.\,H., J.\,S., C.\,M. and S.\,M. fabricated HBL devices. B.\,P., J.\,S., J.\,F. and S.\,M. performed optical spectroscopy. A.\,S.\,B. developed theoretical models and performed numerical calculations.  B.\,P., J.\,S. and A.\,H. analyzed the data and wrote the manuscript. B.\,P. and J.\,S. contributed equally to this work.
%%%%%%%%%%%%%%%%%%%%%%%%%%%%%%%%%%%%%%%%%%%%%%%%%%%%%%%%%%%%%%%%%%%%%%%%%%%%%%%%%%%
\vspace{8pt}
\\
%%%%%%%%%%%%%%%%%%%%%%%%%%%%%%%%%%%%%%%%%%%%%%%%%%%%%%%%%%%%%%%%%%%%%%%%%%%%%%%%%%%
\textbf{Corresponding authors:}\\
B.\,P. (borislav.polovnikov@physik.uni-muenchen.de), J.\,S. (johannes.scherzer@physik.uni-muenchen.de), A.\,B. (anvar.baimuratov@lmu.de), and A.\,H. (alexander.hoegele@lmu.de). 
%%%%%%%%%%%%%%%%%%%%%%%%%%%%%%%%%%%%%%%%%%%%%%%%%%%%%%%%%%%%%%%%%%%%%%%%%%%%%%%%%%%
\vspace{8pt}
\\
%%%%%%%%%%%%%%%%%%%%%%%%%%%%%%%%%%%%%%%%%%%%%%%%%%%%%%%%%%%%%%%%%%%%%%%%%%%%%%%%%%%
\textbf{Data availability:} 
The data that support the findings of this study are available from the corresponding authors upon reasonable request.
%%%%%%%%%%%%%%%%%%%%%%%%%%%%%%%%%%%%%%%%%%%%%%%%%%%%%%%%%%%%%%%%%%%%%%%%%%%%%%%%%%%
\vspace{8pt}
\\
%%%%%%%%%%%%%%%%%%%%%%%%%%%%%%%%%%%%%%%%%%%%%%%%%%%%%%%%%%%%%%%%%%%%%%%%%%%%%%%%%%%
\textbf{Code availability:} 
The codes that support the findings of this study are available from the corresponding authors upon reasonable request.
%%%%%%%%%%%%%%%%%%%%%%%%%%%%%%%%%%%%%%%%%%%%%%%%%%%%%%%%%%%%%%%%%%%%%%%%%%%%%%%%%%%
\vspace{8pt}
\\
%%%%%%%%%%%%%%%%%%%%%%%%%%%%%%%%%%%%%%%%%%%%%%%%%%%%%%%%%%%%%%%%%%%%%%%%%%%%%%%%%%%
\textbf{Competing interests:}\\
The authors declare no competing interests.
%%%%%%%%%%%%%%%%%%%%%%%%%%%%%%%%%%%%%%%%%%%%%%%%%%%%%%%%%%%%%%%%%%%%%%%%%%%%%%%%%%%

%\clearpage

\bibliography{MoSe2WS2_bibliography_v4}

%%%%%%%%%%%%%%%%%%%%%%%%% PLACE FIGS HERE
%%%%%%%%%%%%%%%%%%%%%%%%% PLACE FIGS HERE
\renewcommand{\figurename}{\textbf{Extended Data Figure}}
\setcounter{figure}{0}

%%%%%%%%%%%%%%%%%%%%%%%%%%%%%%%%% Fig. S1 %%%%%%%%%%%%%%%%%%%%%%%%%%%%%%%%%
\begin{figure*}[t]
\centering
\includegraphics[scale=1.0]{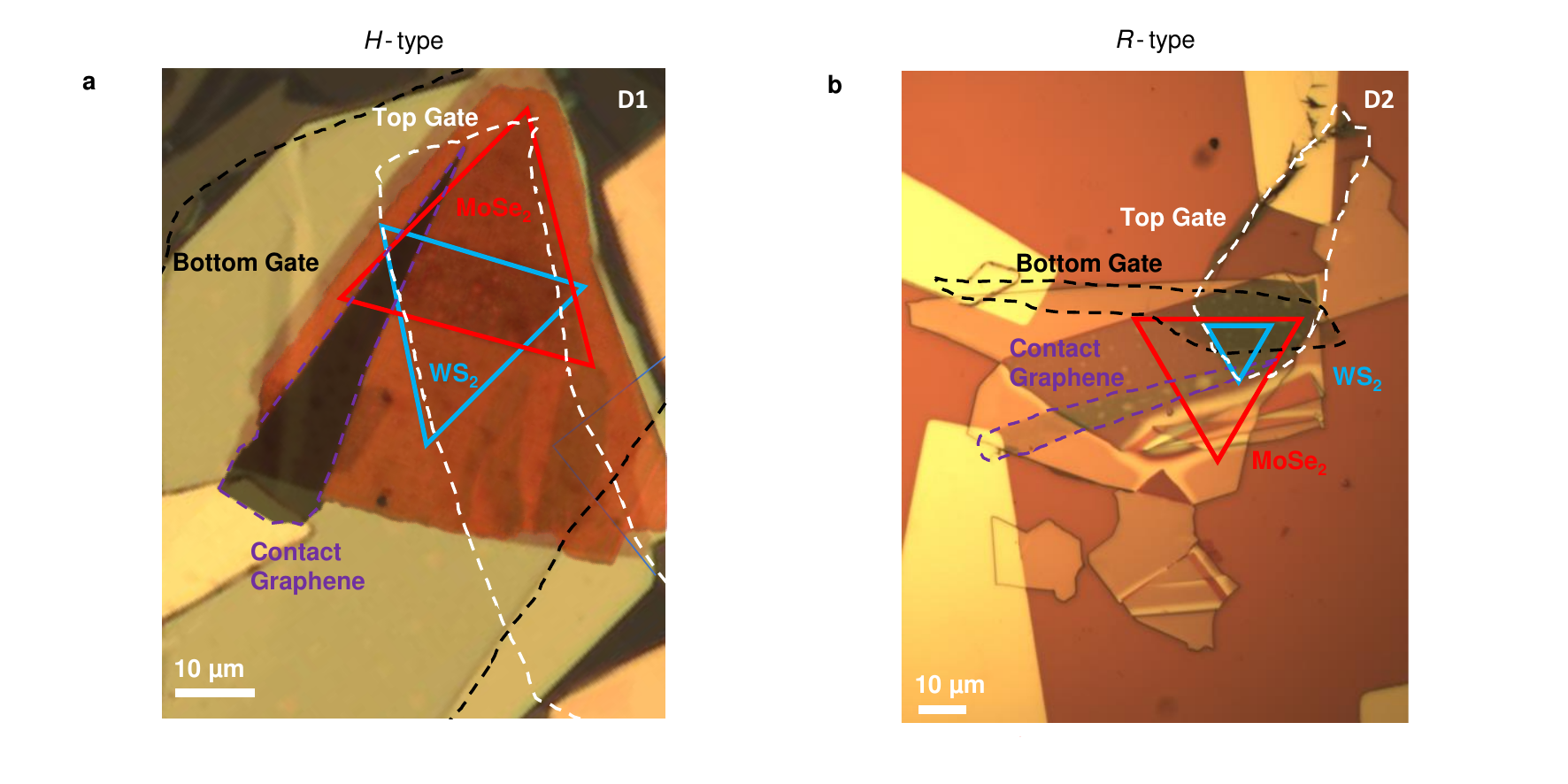}
\caption{\textbf{Layout of devices D1 and D2.} \textbf{a} and \textbf{b}, Optical images of devices D1 and D2, respectively, assembled from CVD-grown triangular single-crystal WS$_2$ and MoSe$_2$ monolayers delimited in red and blue. All other relevant elements of the field-effect structure are framed by dashed lines in respective colors. The data in the main text were acquired in regions where the heterobilayer is sandwiched between both top and bottom gates to ensure symmetric doping and linear scaling of the electric field with voltages applied to the gates.}
\label{SI_fig1}
\end{figure*}
%%%%%%%%%%%%%%%%%%%%%%%%%%%%%%%%%%%%%%%%%%%%%%%%%%%%%%%%%%%%%%%%%%

%%%%%%%%%%%%%%%%%%%%%%%%%%%%%%%%% Fig. S2 %%%%%%%%%%%%%%%%%%%%%%%%%%%%%%%%%
\begin{figure*}[t]
\centering
\includegraphics[scale=1.0]{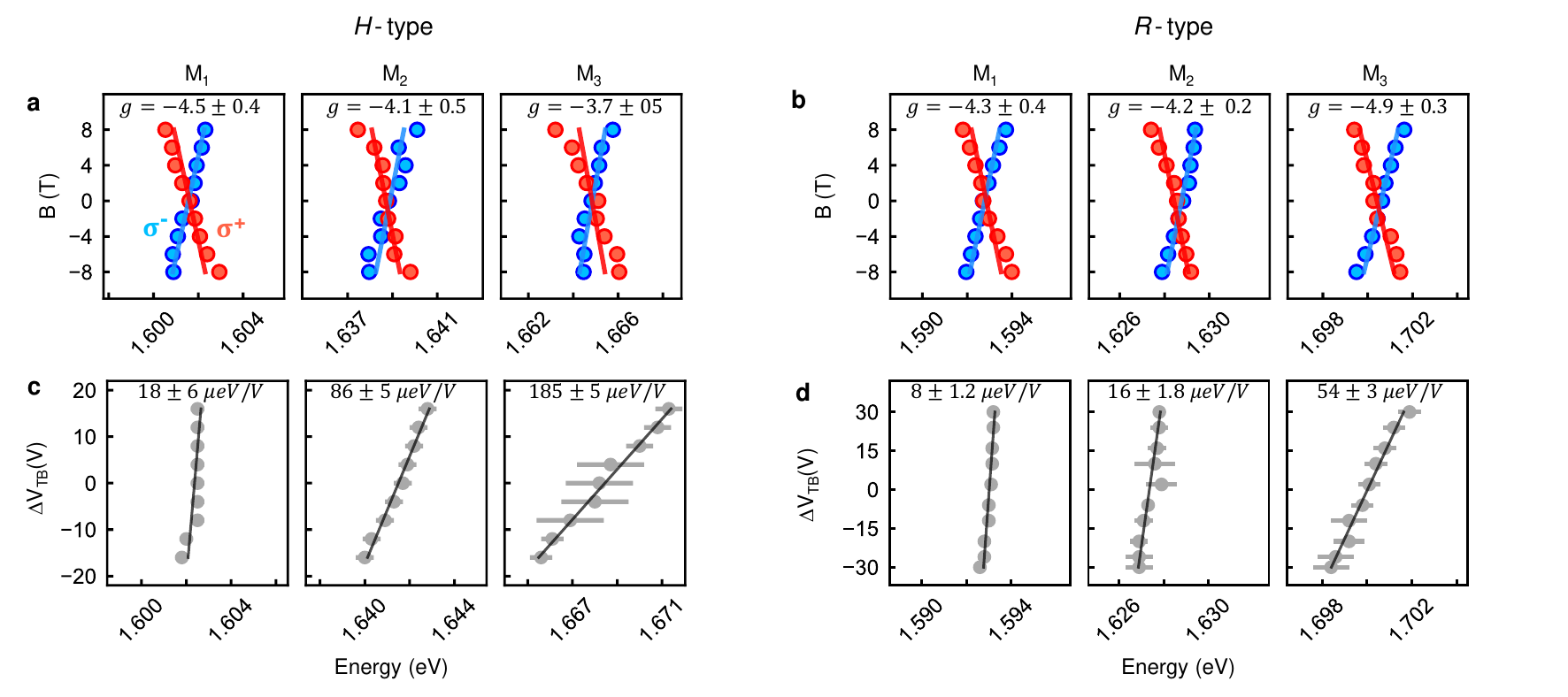}
\caption{\textbf{Dispersions of neutral moir\'e excitons in electric and magnetic fields.} \textbf{a} and \textbf{b}, Dispersions of peaks $M_1$, $M_2$ and $M_3$ in magnetic field for devices D1 and D2 in $H$- and $R$-type stacking, respectively. The $g$-factors were determined from linear fits (red and blue solid lines) to the magneto-dispersions of circularly polarized branches (red and blue data). \textbf{c} and \textbf{d}, Dispersion of $M_1$, $M_2$ and $M_3$ peak energies in electric field applied by imbalanced top and bottom gate voltages in devices D1 and D2 with $H$- and $R$-stacking, respectively. The numbers in each panel indicate the linear slopes of the first-order Stark effect.}
\label{SI_fig2}
\end{figure*} 
%%%%%%%%%%%%%%%%%%%%%%%%%%%%%%%%%%%%%%%%%%%%%%%%%%%%%%%%%%%%%%%%%%

%%%%%%%%%%%%%%%%%%%%%%%%%%%%%%%%% Fig. S3 %%%%%%%%%%%%%%%%%%%%%%%%%%%%%%%%%
\begin{figure*}[t]
\centering
\includegraphics[scale=1.0]{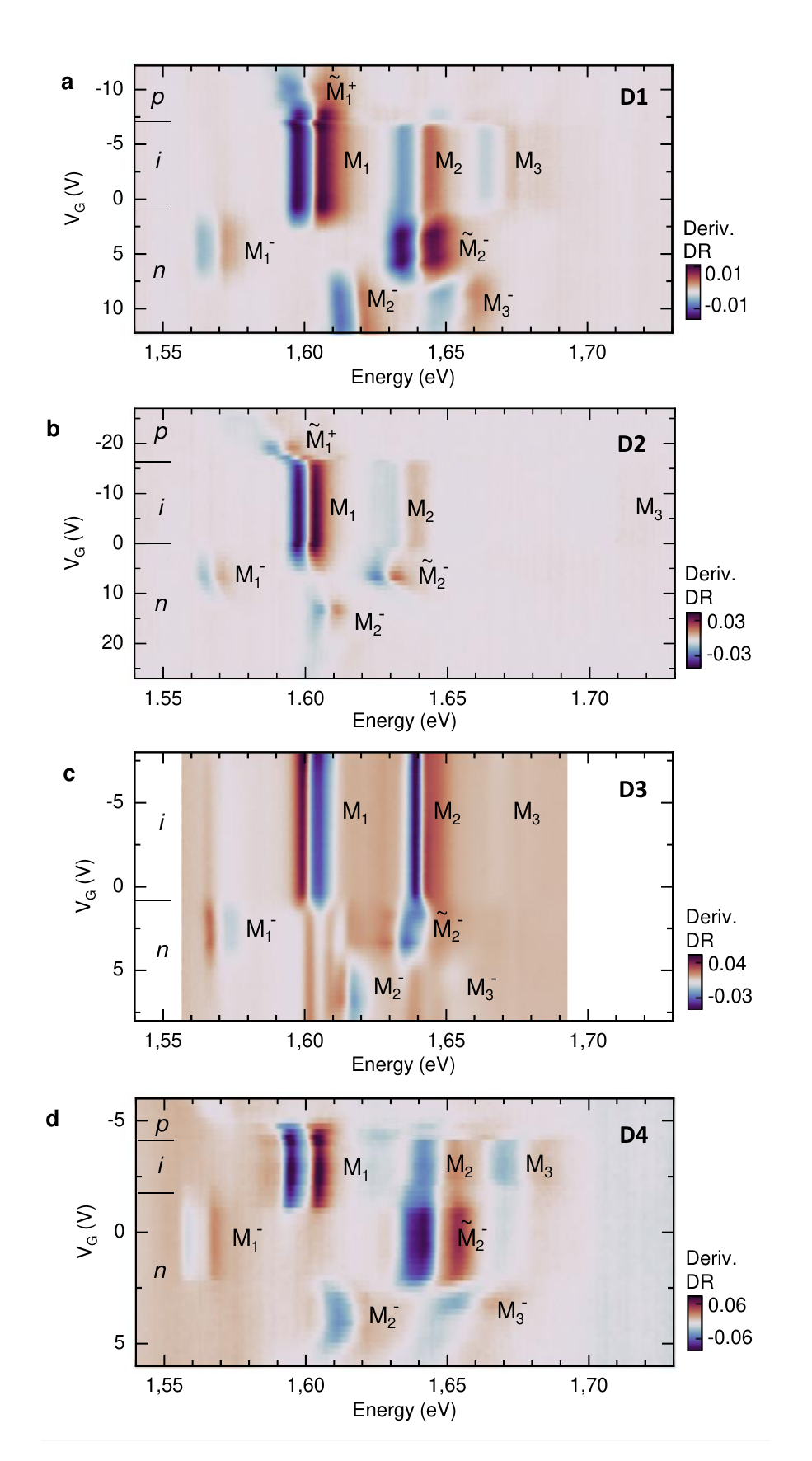}
\caption{\textbf{Ambipolar field-effect doping regimes of MoSe2/WS2 heterostacks}. \textbf{a} - \textbf{d}, Evolution of the derivative of DR with respect to energy for devices D1 and D2 and two additional devices D3 and D4 assembled from monolayers exfoliated from native crystals. The data in \textbf{a} and \textbf{b} show the derivatives of the data in Fig. 3\textbf{c} and \textbf{d} of the main text. The charging diagrams of devices D3 and D4 are consistent with $H$-type stacking.}
\label{SI_fig3}
\end{figure*} 
%%%%%%%%%%%%%%%%%%%%%%%%%%%%%%%%%%%%%%%%%%%%%%%%%%%%%%%%%%%%%%%%%%

%%%%%%%%%%%%%%%%%%%%%%%%%%%%%%%%% Fig. S4 %%%%%%%%%%%%%%%%%%%%%%%%%%%%%%%%%
\begin{figure*}[t]
\centering
\includegraphics[scale=1.0]{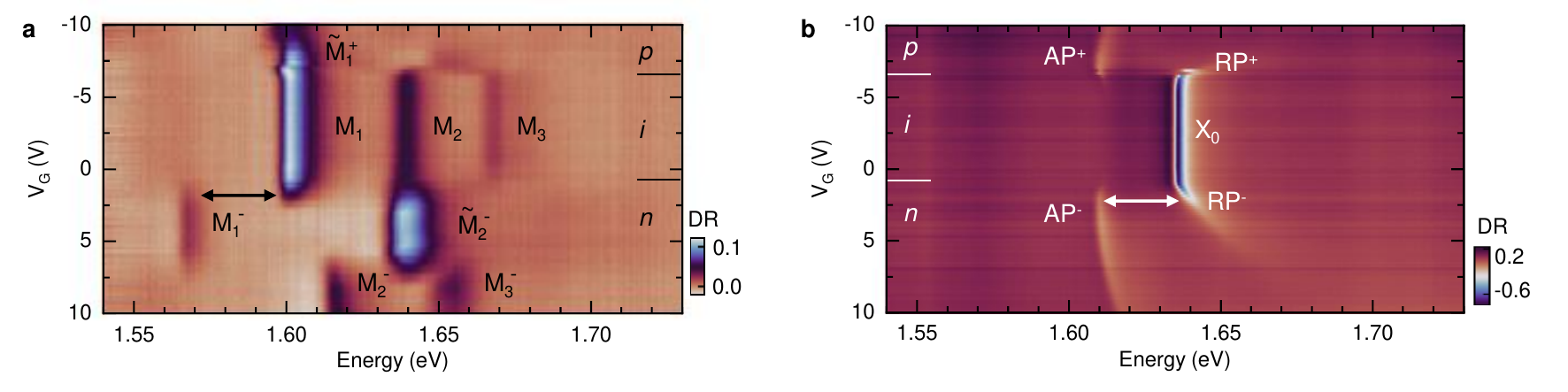}
\caption{\textbf{Comparison of MoSe$_2$/WS$_2$ heterobilayer and MoSe$_2$ monolayer charging diagrams.} \textbf{a} Charging diagram of device D1 as shown in Fig.~3c of the main text. \textbf{b} Corresponding charging diagram recorded on the MoSe$_2$ monolayer region of the same device. Both diagrams feature characteristic signatures of ambipolar doping at the same gate voltages, with positive and negative attractive polaron ($AP$) and repulsive polaron ($RP$) branches in monolayer MoSe$_2$. The energy splitting of $28$~meV between the negative polaron branches ($RP^-$ and $AP^-$) at the onset of electron-doping in monolayer MoSe$_2$ is comparable to the splitting of $33$~meV between the $M_1$ and $M_1^-$ moir\'e peaks as indicated by the arrows.}
\label{SI_fig4}
\end{figure*} 
%%%%%%%%%%%%%%%%%%%%%%%%%%%%%%%%%%%%%%%%%%%%%%%%%%%%%%%%%%%%%%%%%%

%%%%%%%%%%%%%%%%%%%%%%%%%%%%%%%%% Fig. S5 %%%%%%%%%%%%%%%%%%%%%%%%%%%%%%%%%
\begin{figure*}[t]
\centering
\includegraphics[scale=1.0]{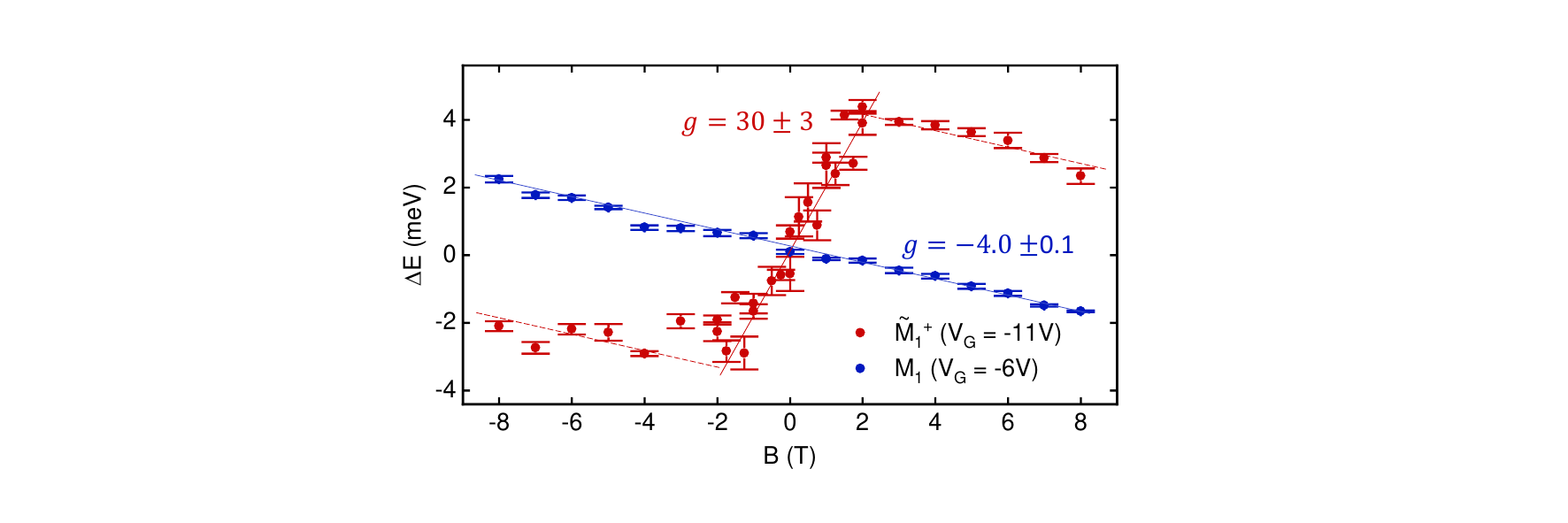}
\caption{\textbf{Evolution of Land\'e factors with magnetic field for the peaks $M_1$ and $\tilde{M}_1^+$.} Evolution of the valley Zeeman splitting $\Delta E$ of the neutral moir\'e peak $M_1$ (blue data and fits) and the hole-doped peak $\tilde{M}_1^+$ (red data and fits) at $V_\text{G}=-11$~V in device D1, obtained as the difference of the peak energies in polarization-resolved DR spectra. The data for $M_1$ were fitted by a linear function with $g=-4.0\pm 0.1$. The Zeeman splitting of $\tilde{M}_1^+$, on the other hand, depends non-linearly on the magnetic field, with a maximum $g$-factor of $30\pm3$ (determined from linear fit for $B\lesssim 2$~T) and an asymptotic $g$-factor of $-4$ at large magnetic fields. The non-linear evolution of the $g$-factor is indicative of correlation-induced magnetism probed by $\tilde{M}_1^+$.}
\label{SI_fig5}
\end{figure*} 
%%%%%%%%%%%%%%%%%%%%%%%%%%%%%%%%%%%%%%%%%%%%%%%%%%%%%%%%%%%%%%%%%%

%
%\setlength{\textheight}{23cm} 
%\setlength{\textwidth}{18cm}
%\setlength{\columnwidth}{18cm}
%\headsep 0mm
%\headheight 0mm
%\setlength{\oddsidemargin}{\dimexpr(\paperwidth-\textwidth)/2-1in}
%\setlength{\evensidemargin}{\oddsidemargin}
%\setlength{\topmargin}{\dimexpr(\paperheight-\textheight)/2-\headheight-\headsep-1in}
%\renewcommand{\baselinestretch}{1.6}
%\renewcommand{\topfraction}{1.0}
%
%\clubpenalty=10000
%\widowpenalty=10000
%\displaywidowpenalty=10000
%
%\setlength{\parindent}{0pt}

\end{document}